\documentclass{webofc}
\usepackage[varg]{txfonts}   
\begin{document}
\title{Finite-temperature equation of state with hyperons }
%
%

\author{\firstname{Hristijan}  \lastname{Kochankovski}\inst{1,2}\fnsep\thanks{\email{hriskoch@fqa.ub.edu}} \and
        \firstname{Angels} \lastname{Ramos}\inst{1}\fnsep\thanks{\email{ramos@fqa.ub.edu}} \and
        \firstname{Laura} \lastname{Tolos}\inst{3,4,5}\fnsep\thanks{\email{tolos@ice.csic.es}}
}

\institute{ Departament de F\'{\i}sica Qu\`antica i Astrof\'{\i}sica and Institut de Ci\`encies del Cosmos, Universitat de Barcelona, Mart\'i i Franqu\`es 1, 08028, Barcelona, Spain
\and
          Faculty of Natural Sciences and Mathematics-Skopje, Ss. Cyril and Methodius University in Skopje Arhimedova, 1000 Skopje, North Macedonia         
 \and
           Institute of Space Sciences (ICE, CSIC), Campus UAB,  Carrer de Can Magrans, 08193 Barcelona, Spain
 \and  
 Institut d' Estudis Espacials de Catalunya (IEEC), 08034 Barcelona, Spain
 \and         
 Frankfurt Institute for Advanced Studies, Ruth-Moufang-Str. 1, 60438 Frankfurt am Main, Germany  
          }

\abstract{%
We present the novel finite-temperature FSU2H$^*$ equation-of-state model that covers a wide range of temperatures and lepton fractions for the conditions in proto-neutron stars, neutron star mergers and supernovae. The temperature effects on the thermodynamical observables and the composition of the neutron star core are stronger when the hyperonic degrees of freedom are considered. We pay a special attention to the temperature and density dependence of the thermal index in the presence of hyperons and conclude that the true thermal effects cannot be reproduced with the use of a constant $\Gamma$ law.}
\maketitle
\section{Introduction}
\label{intro}

Neutron stars are one of the most compact objects in the Universe. Because of their extreme properties, they are a natural laboratory for studying matter under extreme
conditions. Specifically, the core of the neutron star is the most fascinating part as we know very little about its composition, whether only nucleonic matter is present or more exotic constituents can appear, such as hyperons \cite{Tolos:2020aln}.

While the description of the cold neutron star core is given by one-parameter equation of state (EoS) that relates the pressure to (energy)density, a finite temperature treatment is mandatory in order to understand the evolution of a young neutron star, the collapse of supernovae or the merger of a binary system of neutron stars. Therefore, the EoS depends on three parameters, such as temperature ($T$), lepton fraction  ($Y_l$) and baryon density $\rho_B$. In particular, in order to consider all previous astrophysical scenarios, the EoS for the core should cover  $T = 0-100$ MeV, $\rho_B = \rho_0/2 -8\rho_0$\footnote{$\rho_0 \sim 0.16 \ {\rm fm^{-3}}$ is the nuclear saturation density}, $Y_l = 0 - 0.4$ \cite{Oertel:2016bki}. 

For these conditions, the appearance of more exotic components beyond nucleons, such as hyperons, is energetically probable, as the high density at the center of the star and the rapid increase of the nucleon chemical potential makes hyperons energetically more favourable than nucleons. 

In the last decade, a lot of effort has been invested in obtaining EoSs applicable to neutron stars, supernovae and binary mergers. Almost one hundred general purpose EoSs have been produced, with about one third taking into account degrees of freedom beyond nucleons (see reviews of \cite{Oertel:2016bki, Burgio:2021vgk,Typel:2022lcx}). However, those numbers are still far from desirable, especially for the models extended to finite temperature.

With the present paper, we aim at improving this situation by presenting the FSU2H$^*$ model at finite temperature \cite{Kochankovski:2022rid}. This model is based on the FSU2H for the nucleonic and hyperonic core of cold neutron stars \cite{Tolos:2016hhl,Tolos:2017lgv}. The FSU2H$^*$ improves in some of the features of the FSU2H model at zero temperature while extending it to finite temperature, so as to address supernovae and neutron star mergers. Moreover, in the present contribution we discuss the validity of the $\Gamma$ law, that is being extensively used in supernovae and binary-merger simulations to account for finite temperature corrections. 

\section{The FSU2H$^*$ model}
\label{fsu2hst}

The FSU2H$^*$ model at finite temperature is a relativistic mean-field model, that considers matter made of baryons and leptons at a given temperature $T$, baryon  density $\rho_B$ and fixed lepton fraction $Y_l$. Within the relativistic mean-field theory, the interaction between the baryons is modelled through the exchange of different mesons, which includes the contributions from the $\sigma$, $\omega$, $\rho$, $\phi$ and $\sigma^{*}$ mesons (see  details in Ref.~\cite{Kochankovski:2022rid}).

In order to obtain the thermodynamic properties and composition of matter at finite temperature, we need to determine
the Dirac equations for the different baryons and leptons, whereas the field equations of motion for the mesons are obtained from the Euler-Lagrange equations at the mean-field level (see  Ref.~\cite{Kochankovski:2022rid} for the description of the model).

These equations have to be solved simultaneously  with the condition of $\beta$-stability inside the core of neutron stars, which is achieved by imposing the following relations between the chemical potentials of the different species in terms of the chemical potential of electrons ($e$), neutrons ($n$), protons ($p$), and anti-muonic neutrinos ($\nu_{\bar{\mu}}$):
\begin{eqnarray}
\label{eq:beta_chemical_potentials}
&&\mu_{b^0} = \mu_n , \nonumber \\
&&\mu_{b^{-}} = 2\mu_n - \mu_p , \nonumber \\
&&\mu_{b^{+}} = \mu_p ,\nonumber \\
&&\mu_n - \mu_p = \mu_e - \mu_{\nu_e},\nonumber \\
&&\mu_e = \mu_{\mu} + \mu_{\nu_e} +\mu_{\bar{{\nu}}_{\mu}} ,
\end{eqnarray}
where $b^0$, $b^{-}$, $b^{+}$ indicate neutral, negatively charged and positively charged baryons, respectively. Furthermore, $\beta$-stable matter is charge neutral, so that
\begin{eqnarray}
0=\sum_{b,l} q_i \rho_i ,
\end{eqnarray}
where $q_i$ denotes the charge of the particle $i$.
Finally, the baryon and lepton densities are also conserved
\begin{eqnarray}
&&\rho_B = \sum_{b} \rho_b, \nonumber \\\
&&Y_l \cdot \rho_B = \rho_{l}+ \rho_{\nu_l} ,
\end{eqnarray}
where $\rho_B$ is the total baryon density, $Y_l$ is the lepton fraction for a given flavour, and $\rho_{l(\nu_l)}$ the lepton (leptonic neutrino) density. Note that in an extreme neutrino-free case $\mu_{\nu_l}={\mu}_{\bar{\nu}_l}=0$ and the lepton number is no longer conserved. 

Then,  the different thermodynamic variables can be then obtained from the stress-energy momentum tensor. We refer the reader to Ref.~\cite{Kochankovski:2022rid} for the final expressions.

The FSU2H$^*$ parametrization starts from the FSU2H one with nucleons and hyperons from Refs.~\cite{Tolos:2016hhl,Tolos:2017lgv}. This model satisfies the $2M_{\odot}$ constraints and  produces stellar radii below 13 km, while fulfilling the saturation properties of nuclear matter and finite nuclei as well as the constraints
on the high-density nuclear pressure coming from heavy-ion collisions. The FSU2H$^*$ scheme improves, however, some of the features of the FSU2H model. On the one hand, the coupling of the $\sigma$ field to the $\Xi$ baryon is modified so as to reproduce the recent $\Xi$ nuclear potential of $U_{\Xi}=-24$ MeV \cite{Friedman:2021rhu}. On the other hand,  the $\sigma^*$ field has been incorporated in the Lagrangian density in order to have a good description of the $\Lambda \Lambda$ bond energy in $\Lambda$ matter \cite{Takahashi:2001nm,Ahn:2013poa}, whereas  the SU(6) value of the $\phi \Lambda$ coupling is kept.  The parameters of the FSU2H$^*$ model are compiled in Tables 1 and 2 of Ref.~\cite{Kochankovski:2022rid}.

\subsection{FSU2H$^*$ model: mass, radius and tidal deformability}

\begin{figure}
    \centering
     \includegraphics[width=0.4\textwidth]{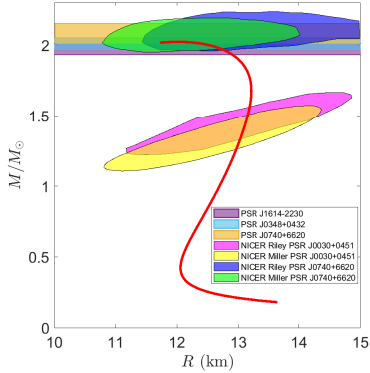}
      \caption{Mass-radius relation for neutron stars coming from the FSU2H$^*$ model, including the constraints from 2$M_{\odot}$ observations \cite{Demorest2010ShapiroStar,Antoniadis:2013pzd,Fonseca2016,NANOGrav:2019jur} and the NICER analyses \cite{Riley:2019yda,Miller:2019cac,Riley:2021pdl,Miller:2021qha}.}
        \label{fig:mass-radius}
\end{figure}

With the FSU2H$^*$ model  the mass and radius of neutron stars can be obtained by solving the structure equations (TOV equations), as similarly done in \cite{Tolos:2016hhl,Tolos:2017lgv} for the FSU2H. The masses and radii obtained with the  FSU2H$^*$  are given in Fig.~\ref{fig:mass-radius} together with the 2$M_{\odot}$ observations \cite{Demorest2010ShapiroStar,Antoniadis:2013pzd,Fonseca2016,NANOGrav:2019jur} and the recent constraints on radii  coming from the two analyses of NICER data \cite{Riley:2019yda,Miller:2019cac,Riley:2021pdl,Miller:2021qha}. Our mass-radius predictions are compatible with all these constraints.

Moreover, we can determine the value of the dimensionless tidal deformability for a 1.4$M_{\odot}$ neutron star and compare it to the results extracted from the gravitational
waves emitted in the binary neutron-star merger GW170817  \cite{TheLIGOScientific:2017qsa,Abbott:2018wiz,Abbott:2018exr}. For the FSU2H$^*$ model a value of 
$\Lambda=526.3$ at M=1.4 $M_\odot$ is obtained, fully compatible with the gravitational wave observations.

\section{Equation of state and composition of hot hyperonic matter}

In order to explore the equation of state and composition of the interior of young neutron stars, remnants of supernova explosions as well as neutron star mergers, we will consider the core of the star at $T = 5$ MeV and $ T = 50$ MeV,  and the neutrino free case as well as neutrino trapped matter with lepton fraction $Y_l = 0.4$. We note that the muon lepton number in all calculations of neutrino trapped matter is fixed to $Y_{\mu}=0$.

\begin{figure}
    \centering
    \includegraphics[width=0.4\textwidth]{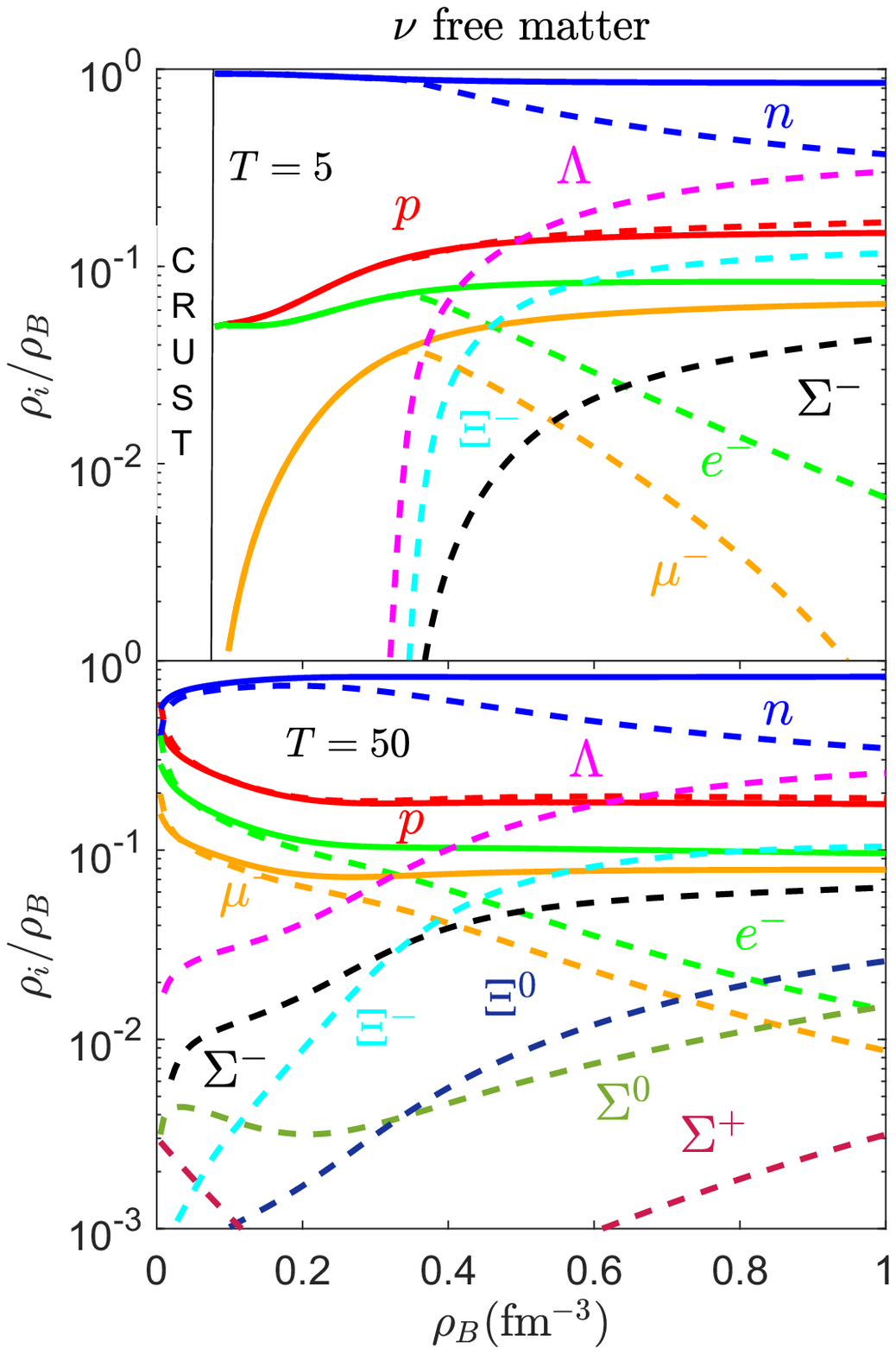}
    \includegraphics[width=0.4\textwidth, height=0.565\textwidth]{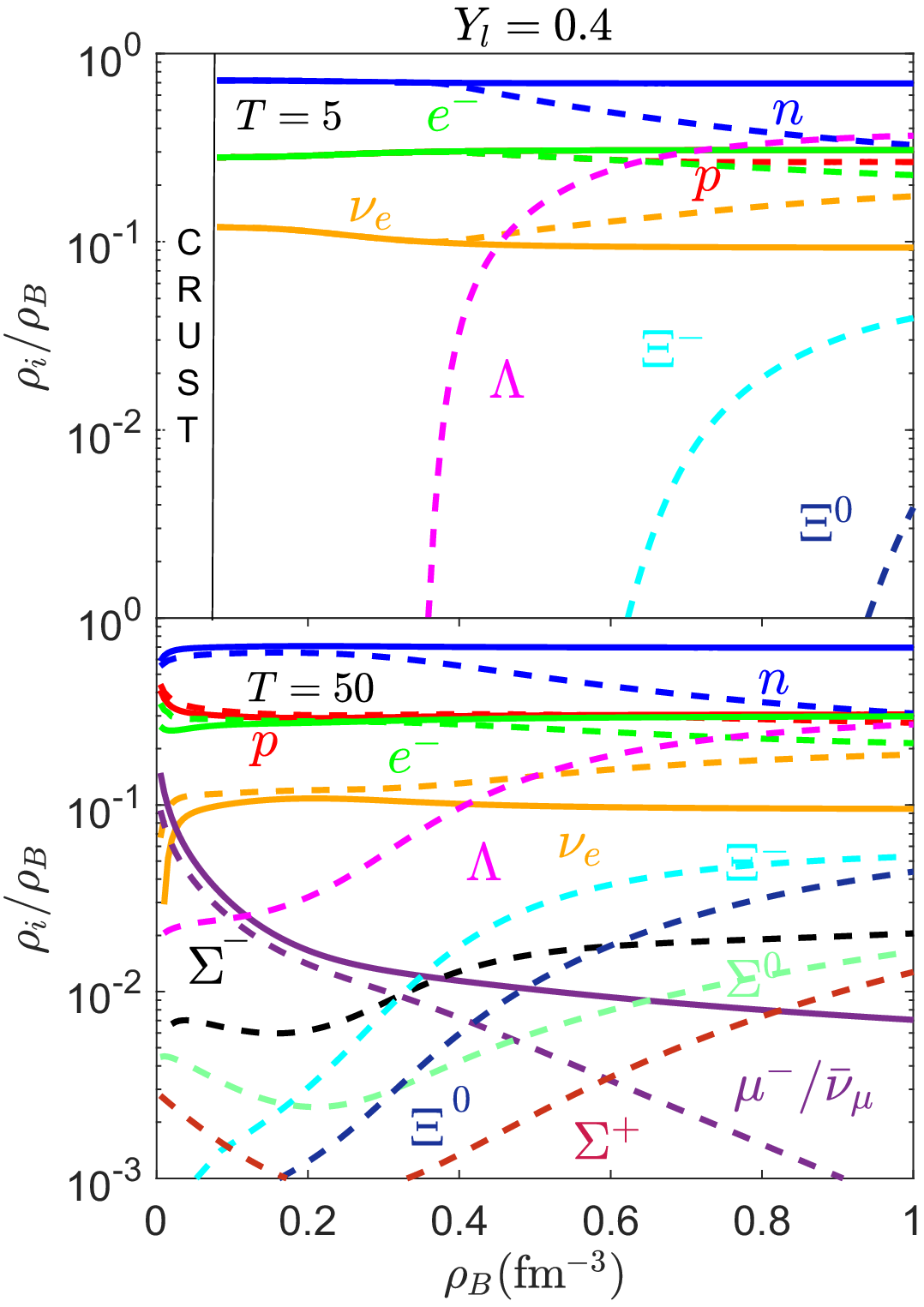}
    \caption{The composition of the neutron star core for neutrino free matter (left panels) and neutrino trapped matter with lepton number $Y_l=0.4$ (right panels), without (solid lines) and with (dashed lines) hyperons, for $T=5$ MeV (upper panels) and $T=50$ MeV (lower panels). Figures taken from Ref.~\cite{Kochankovski:2022rid}.}
    \label{fig:composition}
\end{figure}

In  Fig.~\ref{fig:composition} we show the composition of a neutrino free neutron star core (left panels) and the composition of neutrino trapped matter in the core with lepton number $Y_l=0.4$ (right panels) for temperatures $T=5$ MeV (upper panels) and $T=50$ MeV (lower panels). The different coloured lines represent the relative abundance of the particles as function of the baryonic density $\rho_B$, being the solid lines those corresponding to the case with only nucleons while the dashed lines consider also hyperons. The composition at temperatures as low as $T=5$ MeV is essentially the same as that for cold neutron star matter \cite{Tolos:2016hhl,Tolos:2017lgv}. 

Several conclusions regarding the presence of hyperons can be extracted from this figure. First, we can see that hyperons make matter more isospin symmetric as neutron and proton abundances come closer in both neutrino free and neutrino trapped matter. Second, in both cases,  at sufficiently high temperatures hyperons are found inside the core at any density. Third, in the neutrino free case we can notice that, as soon as hyperons appear, a deleptonization process happens. Negative hyperons ($\Sigma^{-}, \Xi^{-}$) now enter the conserving charge neutrality condition and they can replace the  leptons, which are slowly vanishing from the core as density increases. And, finally, hyperons increase the neutrino abundance when neutrinos are trapped in the core. This is due to the fact that the appearance of negative hyperons lowers the abundance of the electrons. Then, as the electron lepton number is conserved, the abundance of neutrinos increases significantly, specially at high densities.

\begin{figure}
    \centering
   \includegraphics[width=0.4\textwidth]{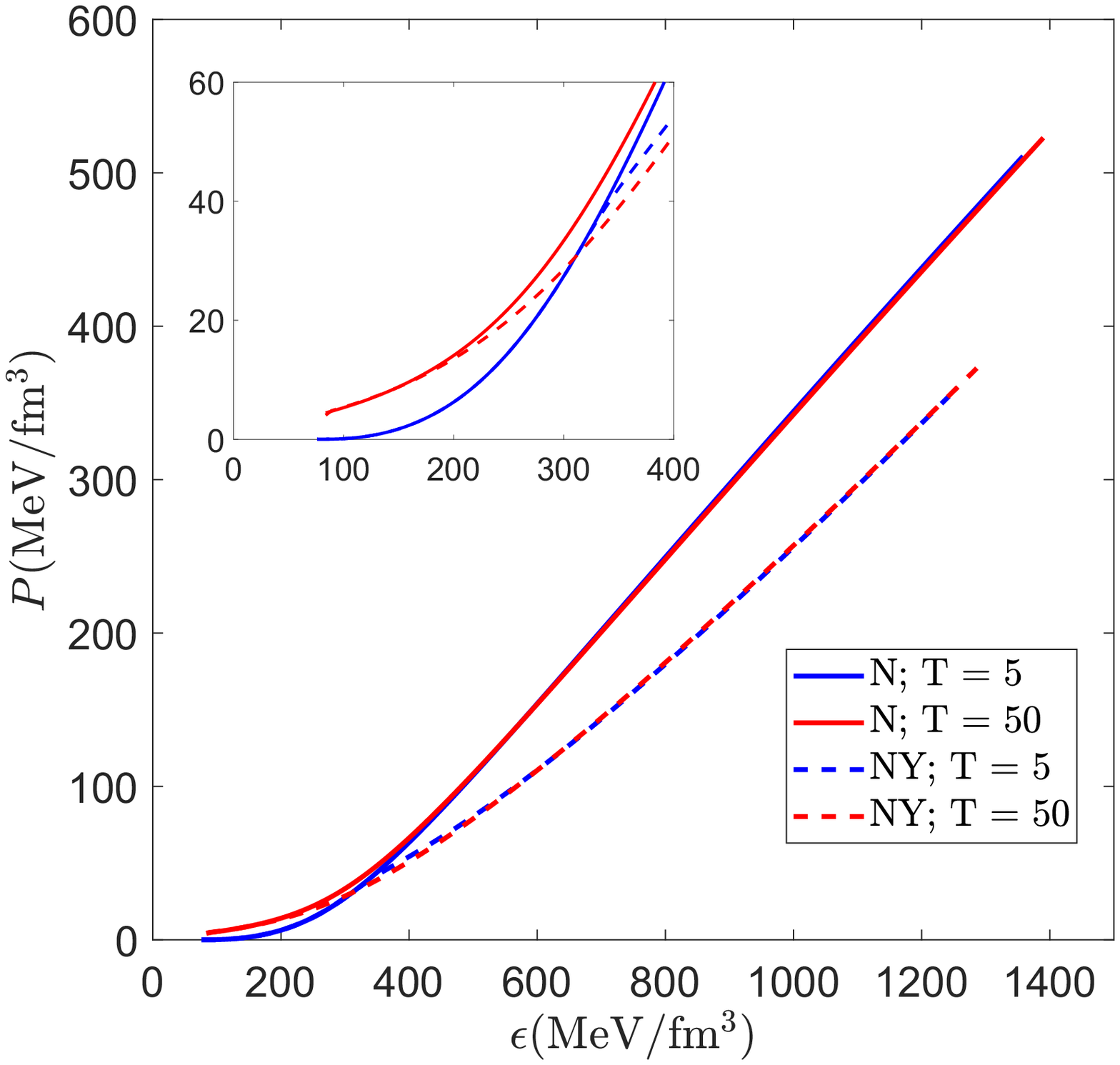}
   \includegraphics[width=0.39\textwidth]{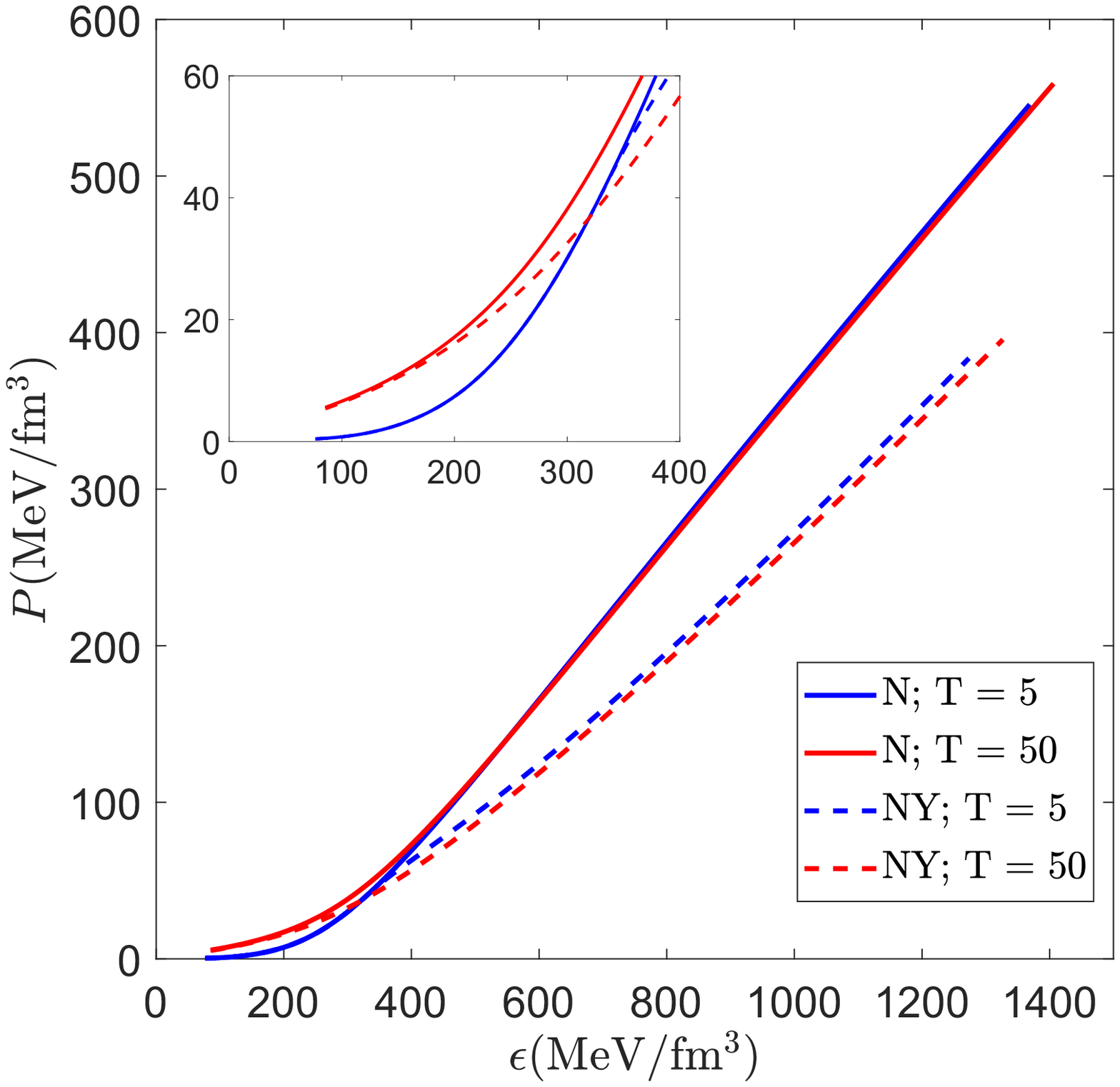}
   \caption{The pressure $P$ as a function of the energy density $\epsilon$ for temperatures $T=5$ MeV and $T=50$ MeV, without (solid lines) and with (dashed lines) hyperons, for the neutrino free case (left panel) and neutrino trapped matter with lepton fraction $Y_l=0.4$ (right panel). Figures modified from Ref.~\cite{Kochankovski:2022rid}.}
    \label{fig:EoS}
\end{figure}

In Fig. ~\ref{fig:EoS} we present the pressure $P$ as a function of the energy density $\epsilon$ for temperatures $T=5$ MeV and $T=50$ MeV, without (solid lines) and with (dashed lines) hyperons, for the neutrino free case (left panel) and neutrino trapped matter with lepton fraction $Y_l=0.4$ (right panel).  The first observation is that hyperons produce a significant softening of the EoS, a very well known fact. Moreover, at low temperatures the production of hyperons drastically changes the slope of the curve, whereas the slope changes more smoothly at high temperatures, since hyperons are already present in the interior of the star at low densities. And, comparing the neutrino free and trapped cases, the EoS becomes stiffer when neutrinos are trapped. In the absence of neutrinos, the appearance of hyperons implies a deleptonization of matter, which in turns makes the EoS softer. In contrast, when the lepton fraction is fixed, the EoS becomes stiffer than the EoS in the neutrino free case since the total density of the leptons remains constant. 

\section{Thermal index of hot hyperonic matter}

We now address the discussion of the thermal index of hot dense matter defined as
\begin{eqnarray}
\centering
    \Gamma(\rho_B,T) \equiv 1 + \frac{P_{\rm th}}{\epsilon_{\rm th}} \ ,
    \label{eq:gamma}
\end{eqnarray}
with the thermal pressure ($P_{\rm th}$) and energy density ($\epsilon_{\rm th}$) given by
\begin{eqnarray}
P_{\rm th}&=& P(\rho_B,T)-P(\rho_B,T=0)  \nonumber \\
\epsilon_{\rm th}&=& \epsilon(\rho_B,T)-\epsilon(\rho_B,T=0)  \ ,
\label{eq:thermal}
\end{eqnarray}
where $P(\rho_B,T=0)$ and $\epsilon(\rho_B,T=0)$ are the pressure and the energy density in cold matter, respectively.

This definition is obtained in analogy with the $\Gamma$ law that relates the pressure and energy density of a diluted ideal gas, where $\Gamma$ is the constant adiabatic index with a value of $5/3 (4/3)$ in the case of non-relativistic (ultra-relativistic) fermionic systems.  In fact, to describe the conditions in supernovae or neutron star mergers while reducing the computational costs, several groups \cite{Hotokezaka:2013iia,Endrizzi:2018uwl,Camelio:2020mdi} use the pressure and energy density of cold matter to which they add the thermal correction, $P_{\rm th}$, that results from the relation established by Eq.~(\ref{eq:gamma}), treating $\epsilon_{\rm th}$ as an independent variable and taking a constant value for $\Gamma$  between 1.5 and 2.0. In this way, an equation of state at finite temperature is calculated. Whereas this approach is fast,  it can be very inaccurate at the conditions of neutron stars mergers.

\begin{figure}
    \centering
    \includegraphics[width=0.4 \textwidth]{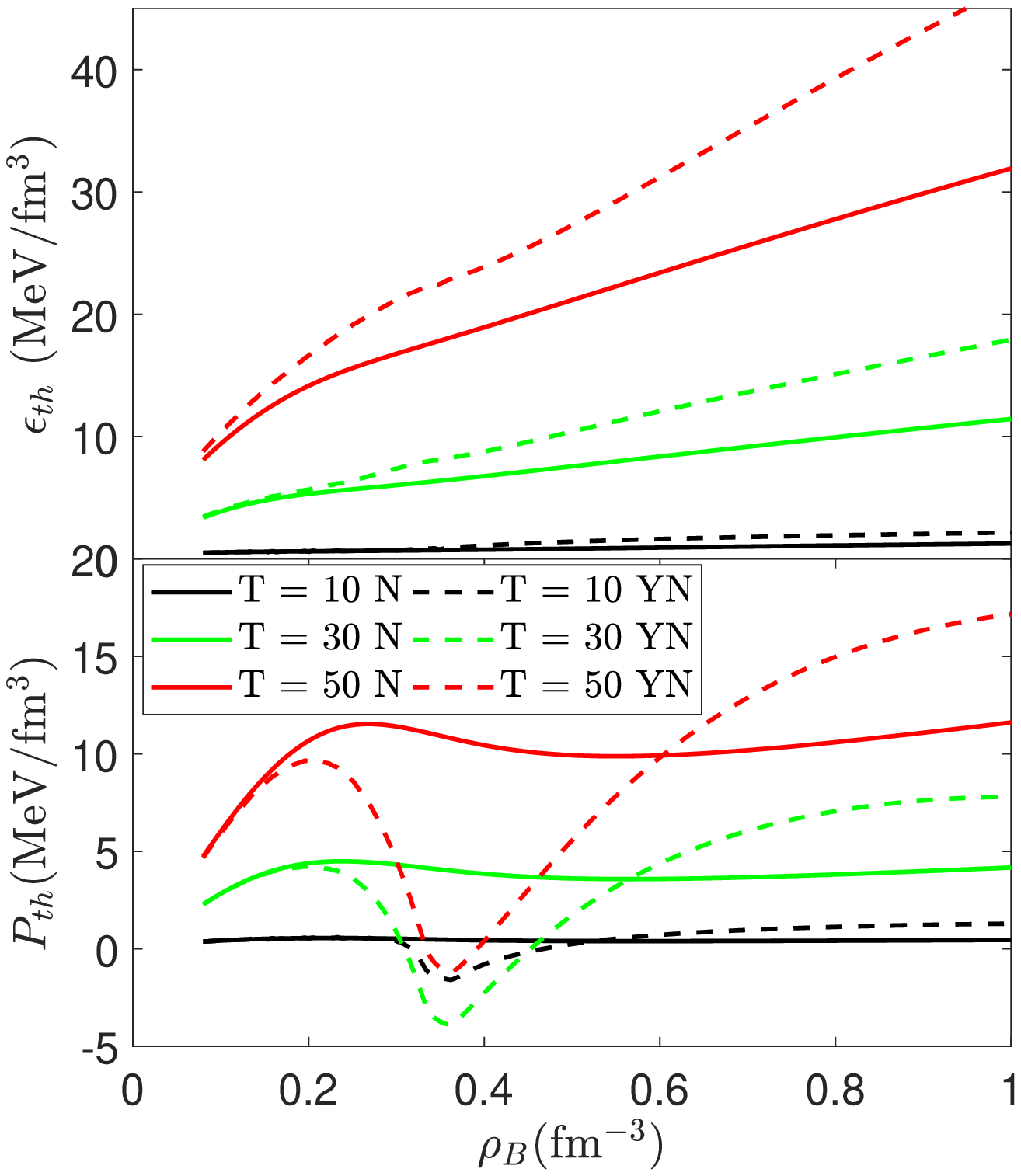}
    \includegraphics[width=0.4 \textwidth]{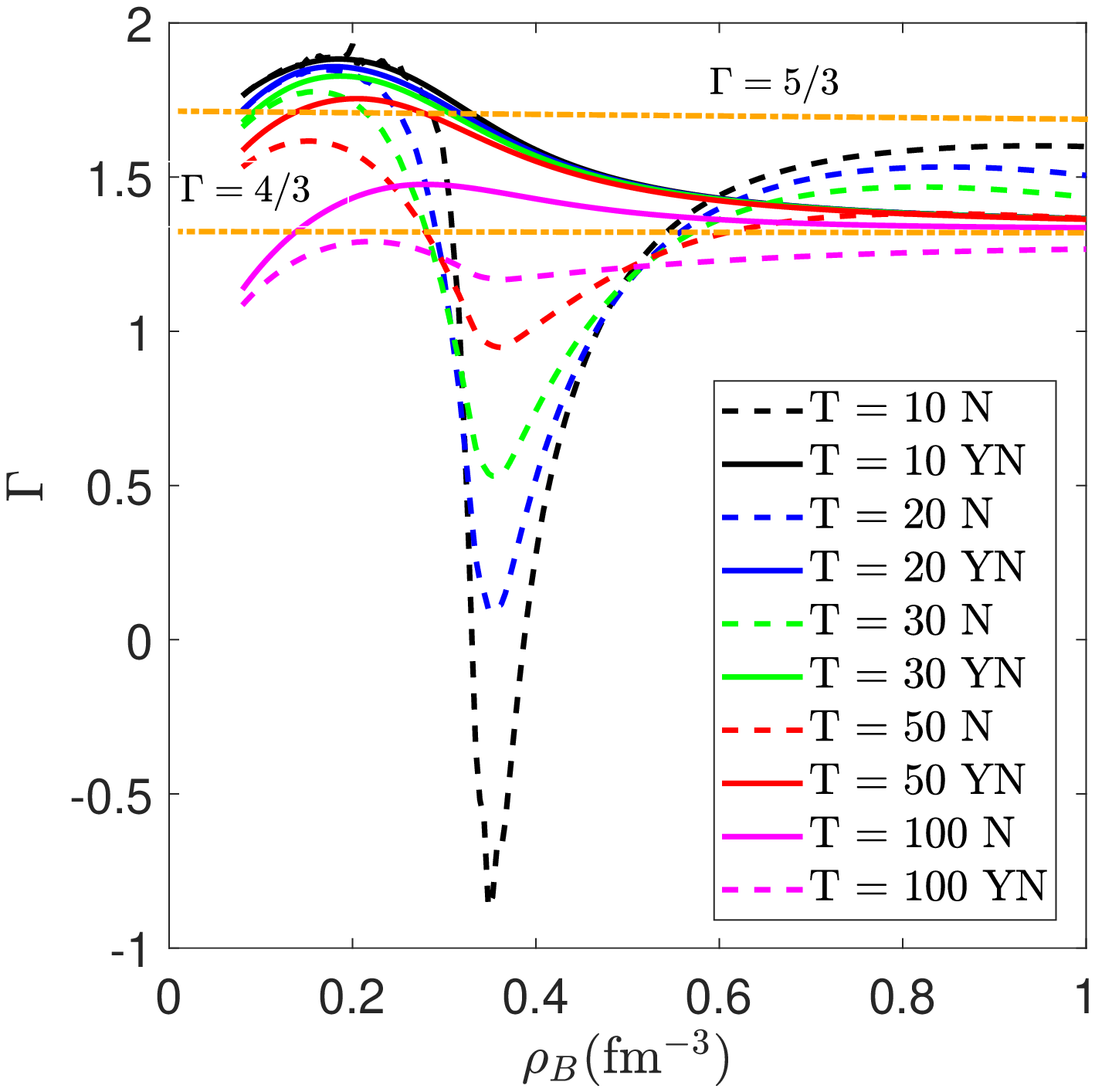}
    \caption{Left panels: The thermal energy $\epsilon_{th}$ and the thermal pressure $P_{th}$ versus the baryonic density $\rho_B$ for different temperatures (in MeV units) and for the neutrino free case. The solid lines correspond to nucleonic matter, whereas the dashed ones are for hyperonic and nucleonic matter.
    Right panel: The thermal index $\Gamma$ versus the baryonic density $\rho_B$ for various temperatures (in MeV units) and for the neutrino free case,  without (solid lines) and with (dashed lines) hyperons. The dashed doted orange lines $\Gamma = 5/3$ and $\Gamma = 4/3$ correspond to the thermal index of a non-relativistic and ultra-relativistic diluted ideal gas, respectively. Figures taken from Ref.~\cite{Kochankovski:2022rid}.}
     \label{fig:thermal}
\end{figure}

In left panels of Fig.~\ref{fig:thermal} we show $P_{\rm th}$ and $\epsilon_{\rm th}$ for three different temperatures as functions of baryon density for nucleonic only matter (solid lines) and hyperonic matter (dashed lines) in the neutrino free case. We observe that the appearance of hyperons has a strong effect on the thermal pressure. Indeed, the thermal pressure  experiences a sizable drop when the hyperon abundance starts being significant. This complex behaviour of the thermal pressure with density heavily influences the thermal index, as seen in the right panel of the same figure. Therefore, the assumption that the thermal index can be mimicked by a constant is inaccurate, specially when hyperons are present.

\begin{figure}
    \centering
    \includegraphics[width=0.46 \textwidth]{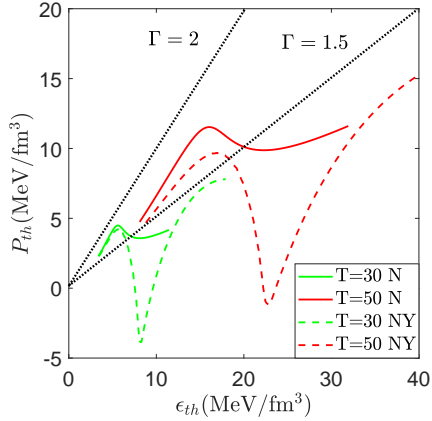}
    \caption{The thermal pressure $P_{th}$ versus the thermal energy density $\epsilon_{th}$ for different temperatures (in MeV units), without (solid lines) and with hyperons (dashed lines). The dotted lines represent $P_{th}(\epsilon_{th})$ for a constant thermal index of $\Gamma=2$ (upper line) and $\Gamma=1.5$ (lower line). Figure taken from Ref.~\cite{Kochankovski:2022rid}.}
    \label{fig:p_thermal_e_thermal}
\end{figure}

 In order to better understand how inaccurate the thermal effects can be when a constant thermal index is assumed,  in Fig.~\ref{fig:p_thermal_e_thermal} we display $P_{\rm th}$ versus $\epsilon_{\rm th}$, employing our nucleonic (solid lines) and hyperonic (dashed lines) equations of state at different temperatures and densities. Each line covers our results for the whole range of density values  ($0.08-1.0$ fm$^{-3}$) at a constant temperature. The dotted black lines represent two typical extreme linear behaviors with $\Gamma=1.5$ and $\Gamma=2$. Our results show that the behaviour of the thermal index is much more complex that the linear assumption, specially when hyperons are considered.

 \section{Summary}
 
In this contribution we have discussed the FSU2H$^*$ model for the hot dense medium in proto-neutron stars, neutron star mergers and supernovae. This model improves  on the hyperonic FSU2H scheme, while extending it to include finite temperature effects. 

The FSU2H$^*$ model at $T=0$ satisfies most important constraints that come from nuclear experiments and astrophysical observations. In particular, the masses and radii predicted are compatible with the 2$M_{\odot}$ observations \cite{Demorest2010ShapiroStar,Antoniadis:2013pzd,Fonseca2016,NANOGrav:2019jur} and the recent constraints on radii  coming from the two analyses of NICER data \cite{Riley:2019yda,Miller:2019cac,Riley:2021pdl,Miller:2021qha}. Also, the tidal deformability for a 1.4$M_{\odot}$ neutron star is in agreement with the gravitational wave observations  \cite{TheLIGOScientific:2017qsa,Abbott:2018wiz,Abbott:2018exr}.

Moreover, we have also investigated the EoS and the composition of neutron star matter at finite temperature with and without hyperons, observing that the thermal corrections have a strong influence on the composition of high dense matter, in particular when hyperons are considered.

The temperature effects have been also analyzed in terms of the thermal index $\Gamma$, which depends non-negligibly on the temperature and density, specially when hyperons are present. Thus, we conclude that thermal effects with $\Gamma$ constant are inaccurate and should be taken with caution in simulations.

\section*{Acknowledgments}
This research has been supported from the projects CEX2019-000918-M, CEX2020-001058-M (Unidades de Excelencia ``Mar\'{\i}a de Maeztu"), PID2019-110165GB-I00 and PID2020-118758GB-I00, financed by the spanish MCIN/ AEI/10.13039/501100011033/, as well as by the EU STRONG-2020 project, under the program  H2020-INFRAIA-2018-1 grant agreement no. 824093, and by PHAROS COST Action CA16214. H.K. acknowledges support from the PRE2020-093558 Doctoral Grant of the spanish MCIN/ AEI/10.13039/501100011033/. L.T. also acknowledges support from the Generalitat Valenciana under contract PROMETEO/2020/023 and from the CRC-TR 211 'Strong-interaction matter under extreme conditions'- project Nr. 315477589 - TRR 211. 
 
\bibliography{references}

\begin{thebibliography}{24}

\bibitem{Tolos:2020aln}
L.~Tolos, L.~Fabbietti, Prog. Part. Nucl. Phys. \textbf{112}, 103770 (2020),
  \texttt{2002.09223}

\bibitem{Oertel:2016bki}
M.~Oertel, M.~Hempel, T.~Kl\"ahn, S.~Typel, Rev. Mod. Phys. \textbf{89}, 015007
  (2017), \texttt{1610.03361}

\bibitem{Burgio:2021vgk}
G.F. Burgio, H.J. Schulze, I.~Vidana, J.B. Wei, Prog. Part. Nucl. Phys.
  \textbf{120}, 103879 (2021), \texttt{2105.03747}

\bibitem{Typel:2022lcx}
S.~Typel et~al., \emph{{CompOSE Reference Manual}} (2022), \texttt{2203.03209}

\bibitem{Kochankovski:2022rid}
H.~Kochankovski, A.~Ramos, L.~Tolos, Mon. Not. Roy. Astron. Soc. \textbf{517},
  507 (2022), \texttt{2206.11266}

\bibitem{Tolos:2016hhl}
L.~Tolos, M.~Centelles, A.~Ramos, Astrophys. J. \textbf{834}, 3 (2017),
  \texttt{1610.00919}

\bibitem{Tolos:2017lgv}
L.~Tolos, M.~Centelles, A.~Ramos, Publ. Astron. Soc. Austral. \textbf{34}, e065
  (2017), \texttt{1708.08681}

\bibitem{Friedman:2021rhu}
E.~Friedman, A.~Gal, Phys. Lett. B \textbf{820}, 136555 (2021),
  \texttt{2104.00421}

\bibitem{Takahashi:2001nm}
H.~Takahashi et~al., Phys. Rev. Lett. \textbf{87}, 212502 (2001)

\bibitem{Ahn:2013poa}
J.K. Ahn et~al. (E373 (KEK-PS)), Phys. Rev. C \textbf{88}, 014003 (2013)

\bibitem{Demorest2010ShapiroStar}
P.~Demorest, T.~Pennucci, S.~Ransom, M.~Roberts, J.~Hessels, Nature
  \textbf{467}, 1081 (2010)

\bibitem{Antoniadis:2013pzd}
J.~Antoniadis et~al., Science \textbf{340}, 6131 (2013), \texttt{1304.6875}

\bibitem{Fonseca2016}
E.~Fonseca, T.T. Pennucci, J.A. Ellis, I.H. Stairs, D.J. Nice, S.M. Ransom,
  P.B. Demorest, Z.~Arzoumanian, K.~Crowter, T.~Dolch et~al., Astrophys. J.
  \textbf{832}, 167 (2016)

\bibitem{NANOGrav:2019jur}
H.T. Cromartie et~al. (NANOGrav), Nature Astron. \textbf{4}, 72 (2019),
  \texttt{1904.06759}

\bibitem{Riley:2019yda}
T.E. Riley et~al., Astrophys. J. Lett. \textbf{887}, L21 (2019),
  \texttt{1912.05702}

\bibitem{Miller:2019cac}
M.C. Miller et~al., Astrophys. J. Lett. \textbf{887}, L24 (2019),
  \texttt{1912.05705}

\bibitem{Riley:2021pdl}
T.E. Riley et~al., Astrophys. J. Lett. \textbf{918}, L27 (2021),
  \texttt{2105.06980}

\bibitem{Miller:2021qha}
M.C. Miller et~al., Astrophys. J. Lett. \textbf{918}, L28 (2021),
  \texttt{2105.06979}

\bibitem{TheLIGOScientific:2017qsa}
B.P. Abbott et~al. (LIGO Scientific, Virgo), Phys. Rev. Lett. \textbf{119},
  161101 (2017), \texttt{1710.05832}

\bibitem{Abbott:2018wiz}
B.P. Abbott et~al. (LIGO Scientific, Virgo), Phys. Rev. \textbf{X9}, 011001
  (2019), \texttt{1805.11579}

\bibitem{Abbott:2018exr}
B.P. Abbott et~al. (LIGO Scientific, Virgo), Phys. Rev. Lett. \textbf{121},
  161101 (2018), \texttt{1805.11581}

\bibitem{Hotokezaka:2013iia}
K.~Hotokezaka, K.~Kiuchi, K.~Kyutoku, T.~Muranushi, Y.i. Sekiguchi, M.~Shibata,
  K.~Taniguchi, Phys. Rev. D \textbf{88}, 044026 (2013), \texttt{1307.5888}

\bibitem{Endrizzi:2018uwl}
A.~Endrizzi, D.~Logoteta, B.~Giacomazzo, I.~Bombaci, W.~Kastaun, R.~Ciolfi,
  Phys. Rev. D \textbf{98}, 043015 (2018), \texttt{1806.09832}

\bibitem{Camelio:2020mdi}
G.~Camelio, T.~Dietrich, S.~Rosswog, B.~Haskell, Phys. Rev. D \textbf{103},
  063014 (2021), \texttt{2011.10557}

\end{thebibliography}

\end{document}